\documentclass[12pt]{article}
\global\parskip 6pt
\newcommand{\bea}{\begin{eqnarray}}
\newcommand{\eea}{\end{eqnarray}}
\newcommand{\non}{\nonumber}

\begin{document}
\begin{titlepage}
\hfill{hep-th/0609028}
\vspace*{.5cm}
\begin{center}
{\Large{{\bf Exact Gravitational Quasinormal Frequencies
of\\[.5ex]
Topological Black Holes}}} \\[.5ex]

\vspace*{1.5cm} Danny Birmingham\footnote{Email: dbirmingham@pacific.edu}\\
\vspace{.1cm}
{\em Department of Physics,\\
University of the Pacific,\\
Stockton, CA 95211\\
USA}\\
\vspace*{1cm} Susan Mokhtari\footnote{Email:
susan@science.csustan.edu}\\
\vspace{.1cm} {\em Department of Physics,\\
California State University Stanislaus,\\
Turlock, CA 95380,\\
USA}
\begin{abstract}
\noindent {We compute the exact gravitational quasinormal
frequencies for massless topological black holes in d-dimensional
anti-de Sitter space. Using the gauge invariant formalism for gravitational
perturbations derived by Kodama and Ishibashi, we show that in all
cases the scalar, vector, and tensor modes can be reduced to a
simple scalar field equation. This equation is exactly solvable in
terms of hypergeometric functions, thus allowing an exact analytic
determination of the gravitational quasinormal frequencies.}
\\
%\vspace*{.25cm}
%\noindent {PACS: 04.70Dy, 04.60.Kz, 11.25.Hf\ \ \ }}
\end{abstract}
\vspace*{.25cm} September 2006
\end{center}
\end{titlepage}

\section{Introduction}

The response of a black hole metric to small perturbations has been
the subject of investigation for many years. By analyzing the perturbation equations
subject to certain boundary conditions, one can gain valuable insight into the
structure of the black hole. Of particular interest is the form of these perturbations
subject to certain quasinormal mode boundary conditions. For the Schwarzschild black hole in four
dimensions, this boundary value problem was analyzed quite some time ago.
It was shown that metric perturbations could be described by either a scalar mode (the Zerilli mode \cite{Z1,Z2})
or a vector mode (the Regge-Wheeler mode \cite{RW}). The resulting wave equations were then solved
approximately (or numerically), leading to a discrete set of complex quasinormal frequencies.
Progress on the analytic solution to the Regge-Wheeler equation has been made recently \cite{Fiziev1,Fiziev2}.
The quasinormal frequencies depend only on the parameters of the black hole, and their
imaginary parts describe the decay of the perturbation in real time.

As a result of the proposed correspondence
between anti-de Sitter gravity and boundary conformal field theory (AdS/CFT) \cite{Maldacena}-\cite{Witten},
quasinormal modes for anti-de Sitter black holes have been investigated intensely, following an earlier analysis
in \cite{ChanMann}.
In particular, it was suggested in \cite{HH}
that the quasinormal modes of a perturbation of an anti-de Sitter black hole
would be related to the timescale for decay of the corresponding operator
in the strongly coupled dual conformal field theory. The return to equilibrium of a perturbed
conformal field theory is well known to be described by linear response theory.
According to the AdS/CFT correspondence, the relaxation time of the conformal field theory
should be related to the quasinormal modes of the black hole perturbation.
Indeed, qualitative evidence was found for this correspondence
by numerically computing the quasinormal modes for scalar field probes in various dimensions \cite{HH}.

A quantitative test of such a correspondence was achieved in \cite{BSS} by analyzing
the quasinormal modes of the three-dimensional BTZ black hole \cite{BTZ}.
The appealing feature of the three-dimensional case is that the quasinormal modes can
be computed exactly \cite{Cardoso1,Bir1,BSS}. Furthermore, on the two-dimensional conformal field theory side,
the retarded correlation functions are also known
explicitly \cite{Keski}. One finds precise quantitative agreement
between the quasinormal
frequencies of the black hole, and the location of the poles (in the momentum representation)  of the
retarded correlation function describing linear response of
the conformal field theory \cite{BSS}.

It is of interest to have examples of black holes in four and higher dimensions
which afford a similar exact analysis. In this regard, we recall that there is a wide
class of topological black holes in anti-de Sitter space, for which the horizon is an Einstein
space of positive, zero, or negative curvature \cite{Lemos}-\cite{Bir2}. In particular, in the negative curvature case,
there is an example of a massless black hole which can play a role quite similar to the
BTZ black hole in three dimensions, and the metric ansatz bears a striking resemblance to
the latter. Indeed, the exact quasinormal modes  of a scalar
field in the background of the massless black hole have been computed \cite{Aros}.
This computation represents the first exact analytic determination of quasinormal modes in four and higher dimensions.
Other works on quasinormal modes of topological black holes have appeared in \cite{Mann2}-\cite{Pap}.

Our goal here is to present an exact analytic computation of the quasinormal modes for all
gravitational perturbations of the massless topological black hole in all dimensions. In order
to achieve this, we make use of the gauge invariant formalism developed recently by Kodama and Ishibashi \cite{Kodama1},
where the equations describing all gravitational perturbations of all higher-dimensional
static black holes have been presented. This powerful formalism identifies three basic types of
gravitational master field, depending on how the field transforms with respect to the horizon manifold.
One has scalar and vector modes, and an additional tensor mode in dimensions greater than four.
Moreover, the equations for these master fields have a standard form as a second order ordinary differential
equation with a potential.
We show that these master field equations for the massless topological black hole
can be solved exactly in all dimensions. The solution is written explicitly in terms of hypergeometric
functions. By imposing appropriate quasinormal mode boundary conditions, we are then led to an exact
determination of all gravitational quasinormal frequencies. We also show that these modes can be
written neatly in a form which gives reference
to the Hawking temperature of the dual conformal field theory, and the conformal weight of
the operator corresponding to the gravitational perturbation.
The bulk computation performed here then gives a prediction for the location of the poles
of the retarded correlation function of the strongly coupled conformal field theory.

The plan of this paper is as follows. In section 2, we recall the essential features of
topological black holes in anti-de Sitter space, and identify the massless topological
black hole with negative curvature horizon. In section 3, we present the basic
equations in the Kodama-Ishibashi formalism for gravitational perturbations. We demonstrate
the unified form which these equations take for the case of the massless topological black hole.
Section 4 is devoted to the explicit solution of the master equations, and the exact
determination of all quasinormal modes.
We conclude in section 5 with a brief discussion of our results within the context of the
AdS/CFT correspondence.

\section{Topological Black Holes in anti-de Sitter Space}
In $d$-dimensional anti-de Sitter space, there is a class of topological black hole solutions
to Einstein's equations which has the property that
the horizon $M^{d-2}$ is a $(d-2)$-dimensional compact Einstein space of positive, zero
or negative curvature $k$ \cite{Lemos}-\cite{Bir2}.
Our interest here is in the negative curvature case $k=-1$.
The line element of the topological black hole is given by \cite{Bir2}
\bea
ds^{2} = - f(r)\; dt^{2} + f^{-1}(r)\;dr^{2} +
r^{2}h_{ij}(x)\;dx^{i}dx^{j},
\label{topbh}
\eea
where
\bea
f(r) = \left(-1 -
\frac{\omega_{d}M}{r^{d-3}} +\frac{r^{2}}{l^{2}}\right),
\eea
and
\bea
\omega_{d} = \frac{16\pi G}{(d-2)\mathrm{Vol}(M^{d-2})}.
\eea
The volume of the horizon is denoted by $\mathrm{Vol}(M^{d-2}) = \int d^{d-2}x\;\sqrt{h}$.
The parameter $l$, with dimensions of length, is related to the cosmological constant $\Lambda$ by
$\Lambda = -(d-1)(d-2)/2l^{2}$, and $\omega_{d}$ is inserted so that $M$ has dimensions of inverse length.

It is straightforward to check that the metric (\ref{topbh}) satisfies Einstein's equations
with negative cosmological constant, namely
\bea
R_{\mu\nu} = -\frac{(d-1)}{l^{2}}g_{\mu\nu}.
\eea
The mass parameter $M$ can be expressed in terms of the location of the horizon $r_{+}$, as
\bea
M = \frac{r_{+}^{d-3}}{\omega_{d}}\left(-1 + \frac{r_{+}^{2}}{l^{2}}\right).
\eea
Furthermore, the inverse Hawking temperature is given by \cite{Bir2}
\bea
\beta = \frac{4\pi l^{2}r_{+}}{(d-1)r_{+}^{2} - (d-3)l^{2}}.
\eea
A very special feature which is present in the case of negative curvature horizon, is that the parameter
$M$ can assume negative values, with the minimal allowed value being
\bea
M_{\mathrm{crit}} = -\left(\frac{2}{d-1}\right)\left(\frac{d-3}{d-1}\right)^{(d-3)/2}\frac{l^{d-3}}{\omega_{d}}.
\eea
The case of interest to us here is the massless topological black hole $M=0$, which has
an event horizon at $r_{+} = l$, and a temperature $T_{H} = 1/2\pi l$. In this case, the function $f$
which is present in the line element assumes the particularly simple form $f(r) = -1 + \frac{r^{2}}{l^{2}}$,
which is reminiscent of the form in the BTZ black hole.
It is also worth noting that the massless black hole is locally isometric to anti-de Sitter
space (i.e., a spacetime of constant curvature) provided that the horizon is itself of constant
curvature \cite{Bir2}.
In this case, the horizon  is a hyperbolic manifold, and is  given as a quotient
$M^{d-2} = H^{d-2}/\Gamma$, where $H^{d-2}$ is hyperbolic space and $\Gamma$ is a suitable discrete
subgroup of the isometry group of $H^{d-2}$.
While these topological black holes are interesting structures in their
own right, they have an important application to the AdS/CFT correspondence. They allow us to study
the dual conformal field theory on spaces of the form $S^{1} \times M^{d-2}$, where
$M^{d-2}$ is an Einstein space of positive, zero, or negative curvature.
Our aim here is to compute the quasinormal modes on the gravity side, and then show how they can be recast
in terms of conformal field theory data.

\section{Gravitational Perturbations}
In order to determine the quasinormal modes of gravitational perturbations of the black hole, we first need to
obtain the relevant equations which describe these perturbations. In the four-dimensional
asymptotically flat case, this was achieved quite some time ago, resulting in the Zerilli equation \cite{Z1,Z2}
and the Regge-Wheeler equation \cite{RW}. These equations were generalized to the anti-de Sitter case in \cite{Cardoso2,
Cardoso3}.
However, a general analysis of gravitational perturbations in higher dimensions was
presented only recently by Kodama and Ishibashi \cite{Kodama1}.
The formalism developed by Kodama and Ishibashi is both powerful and elegant, and is based on
the introduction of gauge invariant variables. These gauge invariant combinations are
then described by master fields $\Phi$. In general,
there are three types of gravitational perturbation; the scalar mode which is the analogue of the Zerilli mode
in higher dimensions, the vector mode which is the analogue of the Regge-Wheeler mode, and an additional
tensor mode which is present in dimensions greater than four. As shown in \cite{Kodama1}, each perturbation
is simply described in terms of a master field $\Phi$ which satisfies a second order  ordinary differential equation
with a potential. These equations have been used to successfully establish the stability
of asymptotically flat Schwarzschild black holes in all dimensions \cite{Kodama2}; an earlier
analysis appeared in \cite{Gibbons}.

To begin, we write the master field as
\bea
\Phi(t,r,x^{i}) = \Phi(r)Y(x^{i})e^{-i\omega t}.
\label{ansatz}
\eea
The type of perturbation then depends on whether $Y$ transforms as a scalar, vector, or tensor
with respect to the horizon manifold $M^{d-2}$. In all cases, however, the
master equation takes the simple form
\bea
\left[-\left(f\frac{d}{dr}\right)^{2} + V - \omega^{2}\right]\Phi(r) = 0,
\label{master}
\eea
where the structure of the potential $V$ depends on the gravitational mode
under consideration. For the scalar mode, we have
\bea
V_{\mathrm{S}}(r) = \frac{f}{r^{2}}\frac{U(r)}{16[\mu + \frac{1}{2}(d-2)(d-1)x]^{2}},
\eea
where
\bea
x = \frac{\omega_{d} M}{r^{d-3}},\;\;\mu = k_{\mathrm{S}}^{2} + (d-2).
\eea
In this case, $Y$ transforms as a scalar, and is an eigenfunction of the scalar Laplacian
$\nabla^{2}Y = -k_{\mathrm{S}}^{2}Y$.
The function $U(r)$ is given by
\bea
U(r) &=& [(d-2)^{3}d(d-1)^{2}x^{2} - 12(d-2)^{2}(d-1)(d-4)\mu x
+ 4(d-4)(d-6)\mu^{2}]\frac{r^{2}}{l^{2}}\non\\
&+& (d-2)^{4}(d-1)^{2}x^{3} + (d-2)(d-1)[4(2(d-2)^{2} - 3 (d-2) + 4)\mu\non\\
&-& (d-2)(d-4)(d-6)(d-1)]x^{2}
- 12(d-2)[(d-6)\mu \non\\
&-& (d-2)(d-1)(d-4)]\mu x
+ 16\mu^{3} -4(d-2)d\mu^{2}.
\eea
The vector mode is described by the potential
\bea
V_{\mathrm{V}}(r) = \frac{f}{r^{2}}\left[k_{\mathrm{V}}^{2} - 1 -
\frac{(d-2)(d-4)}{4} + \frac{(d-2)(d-4)}{4}\frac{r^{2}}{l^{2}} - \frac{3 (d-2)^{2}\omega_{d}M}{4r^{d-3}}\right],
\eea
where $\nabla^{2} Y  = -k_{\mathrm{V}}^{2}Y$.
Finally, the tensor mode in dimension $d>4$ has the potential
\bea
V_{\mathrm{T}}(r) = \frac{f}{r^{2}}\left[k_{\mathrm{T}}^{2} - 2 -
\frac{(d-2)(d-4)}{4} + \frac{d(d-2)}{4}\frac{r^{2}}{l^{2}} + \frac{ (d-2)^{2}\omega_{d}M}{4r^{d-3}}\right],
\eea
with $\nabla^{2} Y  = -k_{\mathrm{T}}^{2}Y$.

The form of these potentials simplifies considerably for the massless topological black hole $M=0$.
The scalar potential is
\bea
V_{\mathrm{S}} = \frac{f}{r^{2}}\left[Q_{\mathrm{S}}
- \frac{(d-2)(d-4)}{4} + \frac{(d-4)(d-6)}{4}\frac{r^{2}}{l^{2}}\right],
\label{Vscalar}
\eea
where we have introduced the notation $Q_{\mathrm{S}} = k_{\mathrm{S}}^{2}$.
The vector potential is
\bea
V_{\mathrm{V}} = \frac{f}{r^{2}}\left[Q_{\mathrm{V}}
- \frac{(d-2)(d-4)}{4} + \frac{(d-2)(d-4)}{4}\frac{r^{2}}{l^{2}}\right],
\label{Vvector}
\eea
with $Q_{\mathrm{V}} = k_{\mathrm{V}}^{2} -1$.
The tensor potential is
\bea
V_{\mathrm{T}} = \frac{f}{r^{2}}\left[Q_{\mathrm{T}} - \frac{(d-2)(d-4)}{4} + \frac{d(d-2)}{
4}\frac{r^{2}}{l^{2}}\right],
\label{Vtensor}
\eea
with $Q_{\mathrm{T}} = k_{\mathrm{T}}^{2} -2$.

Before solving the above equations, let us first examine the case of a scalar
field $\phi$ of mass $m$ in the background of the massless black hole.  The equation of motion for the scalar
field is
\bea
(\nabla^{2} - m^{2})\phi =0.
\eea
Choosing the ansatz
\bea
\phi = \phi(r)Y(x^{i})e^{-i\omega t},
\eea
brings the radial equation to the form (\ref{master}), where $\Phi = r^{\frac{d-2}{2}}\phi$.
The potential is given by
\bea
V = \frac{f}{r^{2}}\left[Q
+f^{\prime}\left(\frac{d-2}{2}\right)r +f \frac{(d-2)(d-4)}{4} + m^{2}r^{2}\right],
\label{Vfield1}
\eea
where $\nabla^{2}Y = -QY$.
Since the metric involves the function $f = -1 + \frac{r^{2}}{l^{2}}$, the potential of the scalar field
takes the particularly simple form
\bea
V = \frac{f}{r^{2}}\left[Q
- \frac{(d-2)(d-4)}{4} + \left(\frac{d(d-2)}{4} + m^{2}l^{2}\right)\frac{r^{2}}{l^{2}}\right].
\label{Vfield}
\eea
In \cite{Aros}, this equation was shown to be exactly solvable in terms of hypergeometric
functions. As a result, the quasinormal mode spectrum of the scalar field could be determined exactly.

We now observe that the gravitational potentials (\ref{Vscalar})-(\ref{Vtensor})
have precisely the same structure as the potential of the scalar field (\ref{Vfield}),
for various values of the mass parameter. We have
\bea
{\mathrm{Scalar\;\; Mode}}:\;\;\;\;\;\; m^{2} l^{2} &=& -2(d-3),\non\\
{\mathrm{ Vector\;\; Mode}}:\;\;\;\;\;\; m^{2}l^{2} &=& -(d-2),\non\\
{\mathrm {Tensor \;\;Mode}}:\;\;\;\;\;\; m^{2} l^{2} &=& 0,
\label{masses}
\eea
with the value $Q$ replaced by the appropriate value $Q_{\mathrm{S}}, Q_{\mathrm{V}}, Q_{\mathrm{T}}$.
It should be noted that the simplicity of the potentials in this case is essentially due to the fact
that the mass parameter of the black hole is set to zero.

\section{Exact Gravitational Quasinormal Modes}

Our aim now is to solve eqn. (\ref{master}), with potentials (\ref{Vscalar})-(\ref{Vtensor}),
subject to quasinormal mode boundary conditions.
In the asymptotically flat case, the quasinormal mode solution is required to be ingoing
at the horizon, and outgoing at asymptotic infinity. As is well know, the wave equation,
subject to these boundary conditions, admits only a discrete set of solutions with complex frequencies.
These quasinormal frequencies have negative imaginary part, thus leading to a decay of the perturbation
via eqn. (\ref{ansatz}). The appropriate boundary conditions for the anti-de Sitter case have been
the subject of recent attention. Firstly, the solution is required to be ingoing at the horizon.
In \cite{HH}, a Dirichlet boundary condition was imposed at infinity, and a numerical determination
of quasinormal modes was presented for scalar fields in various black hole backgrounds.
However, in \cite{BSS}, a subtlety was uncovered due to the allowed
negative mass spectrum of a scalar field in anti-de Sitter space \cite{BF1,BF2}. It was found that vanishing flux
was the more appropriate condition to impose on the scalar field at infinity.
Here, we are concerned with the solution to the master equation for the scalar, vector, and tensor
gravitational modes. We demand the solution to be ingoing at the horizon, and we impose a Dirichlet
boundary condition at infinity.
We shall find that these boundary conditions lead to an exact determination of the gravitational
quasinormal
modes in all dimensions. It should be noted that the
Dirichlet boundary condition at infinity is indeed consistent with the AdS/CFT
correspondence \cite{Keski,Starinets1,Kovtun}. Gravitational quasinormal modes of higher-dimensional black holes
have also been discussed in \cite{Konoplya,Schiappa}.

To proceed towards the solution of (\ref{master}), we change variables to
\bea
z = 1 - \frac{l^{2}}{r^{2}}.
\eea
Thus, $z=0$ corresponds to the location of the horizon $r = l$, while $z=1$ corresponds to
$r=\infty$.
The master equation then becomes
\bea
z(1-z)\frac{d^{2}\Phi}{dz^{2}} + \left(1 - \frac{3z}{2}\right)\frac{d\Phi}{dz}
+\left[\frac{A}{z} + B + \frac{C}{1-z}\right]\Phi = 0,
\eea
where
\bea
A &=& \frac{\omega^{2}l^{2}}{4}, \non\\
B &=& \frac{1}{4}\left(\frac{(d-2)(d-4)}{4} - Q\right), \non\\
C &=& -\frac{1}{4}\left(m^{2}l^{2} + \frac{d(d-2)}{4}\right).
\eea
We now define
\bea
\Phi(z) = z^{\alpha}(1-z)^{\beta}F(z).
\eea
The master equation then reduces to the standard form of the hypergeometric equation
\bea
z(1-z)\frac{d^{2}F}{dz^{2}} + [c - (a+b+1)z]\frac{dF}{dz}
- ab F = 0,
\label{hyper}
\eea
provided that
\bea
\alpha &=& \pm\frac{i\omega l}{2},\non\\
\beta &=& \frac{1}{4} \pm \frac{1}{4}\sqrt{(d-1)^{2} + 4m^{2}l^{2}},
\label{beta}
\eea
with the coefficients determined as followed
\bea
a &=& \frac{1}{4} + \alpha + \beta  + \frac{i\xi}{2}, \non\\
b &=& \frac{1}{4} + \alpha + \beta - \frac{i\xi}{2},\non\\
c &=& 2 \alpha + 1.
\label{abc}
\eea
Here, we have defined $\xi^{2} = Q - \left(\frac{d-3}{2}\right)^{2}$. As noted in \cite{Aros},
$\xi$ is real and continuous in general (at least for the case of the scalar mode).
However, if one considers the case of a hyperbolic horizon, then
the allowed values of $\xi$ typically become discrete.
Without loss of generality, we can take
\bea
\alpha &=&  -\frac{i\omega l}{2}, \non\\
\beta &=& \frac{1}{4} - \frac{1}{4}\sqrt{(d-1)^{2} + 4m^{2}l^{2}}.
\label{alphabeta}
\eea

In the neighbourhood of the horizon, the two linearly independent solutions of (\ref{hyper}) are
$F(a,b,c,z)$ and $z^{1-c}F(a-c+1, b-c+1,2-c,z)$. With the choice (\ref{alphabeta}),
the solution which is ingoing at the horizon is then given by
\bea
\Phi(z) = z^{\alpha}(1-z)^{\beta}F(a,b,c,z).
\eea
Having imposed the ingoing condition at the horizon, we can now analytically continue
this solution to infinity. In general, the form of the solution near $z=1$ is given by \cite{Abram}
\bea
\Phi &=& z^{\alpha}(1-z)^{\beta}\frac{\Gamma(c)\Gamma(c-a-b)}{\Gamma(c-a)\Gamma(c-b)}F(a,b,a+b-c+1,1-z)\non\\
&+& z^{\alpha}(1-z)^{\beta + c-a-b}\frac{\Gamma(c)\Gamma(a+b-c)}{\Gamma(a)\Gamma(b)}F(c-a, c-b, c-a-b+1, 1-z).
\label{infty}
\eea
However, special care is needed when $c-a-b$ is an integer. Therefore, we should examine the coefficients closely,
case by case.
As we have seen, the gravitational scalar mode corresponds to a scalar field with mass $m^{2}l^{2} = -2(d-3)$,
the gravitational vector mode corresponds to a scalar field of mass $m^{2} l^{2} = -(d-2)$, and the
gravitational tensor mode
corresponds to a massless scalar field.
It will be useful to record the values of the coefficient $\beta$ given by (\ref{alphabeta}) for the three
gravitational modes, as follows:
\bea
\beta_{S} &=&
\left\{
  \begin{array}{ll}
    0, & d=4, \\
    -\left(\frac{d-6}{4}\right), & d\geq 5,
  \end{array}
\right.\non\\
\beta_{V} &=& -\left(\frac{d-4}{4}\right),\;\; d\geq 4,\non\\
\beta_{T} &=& -\left(\frac{d-2}{4}\right),\;\;d\geq 4,
\label{beta2}
\eea
where the subscript on $\beta$ specifies the particular mode. From (\ref{abc}), we also note that
$c-a-b = \frac{1}{2} - 2\beta$.

Let us consider first the case in four dimensions. The scalar and vector modes both
have a value of $\beta = 0$, and there is no tensor mode in four dimensions.
Here, $c-a-b = 1/2$, so the analytic continuation to $z=1$ is given by (\ref{infty}). The master field
then takes the form
\bea
\Phi &=& z^{\alpha}\frac{\Gamma(c)\Gamma(c-a-b)}{\Gamma(c-a)\Gamma(c-b)}F(a,b,a+b-c+1,1-z)\non\\
&+& z^{\alpha}(1-z)^{1/2}\frac{\Gamma(c)\Gamma(a+b-c)}{\Gamma(a)\Gamma(b)}F(c-a, c-b, c-a-b+1, 1-z).
\eea
The second term above clearly vanishes at infinity. By imposing Dirichlet boundary conditions on the
perturbation, $\Phi =0$ at $z=1$, we see that the quasinormal modes are determined by the location of the poles
of the Gamma function, as
\bea
c-a = -n, \;\; {\mathrm{or}}\;\; c-b = -n,
\label{qnm}
\eea
where $(n=0,1,2,3,...)$.
In terms of the parameters of the black hole, the scalar quasinormal modes
take the form
\bea
\omega_{\mathrm{S}} = \pm\frac{\xi_{\mathrm{S}}}{l} -\frac{2i}{l}\left( n + \frac{3}{4}\right),
\label{qnm4d}
\eea
where $\xi_{\mathrm{S}}^{2} = Q_{\mathrm{S}} - \frac{1}{4}$. A similar expression holds for
the vector modes, with $\xi_{S}$ replaced by $\xi_{V}$. In the asymptotically flat case, it is known
that the scalar (Zerilli) and vector (Regge-Wheeler)
quasinormal modes are identical for the Schwarzschild black hole
in four dimensions. This is not the case for anti-de Sitter gravity.

Next, we consider all even dimensions greater than four. From (\ref{beta2}), we see that
$\beta \leq 0$ for all perturbations;  furthermore, $c-a-b$ is not an integer.
Thus, the continuation to $z=1$ is given by (\ref{infty})
\bea
\Phi &=& z^{\alpha}(1-z)^{\beta}\frac{\Gamma(c)\Gamma(c-a-b)}{\Gamma(c-a)\Gamma(c-b)}F(a,b,a+b-c+1,1-z)\non\\
&+& z^{\alpha}(1-z)^{\frac{1}{2} - \beta}\frac{\Gamma(c)\Gamma(a+b-c)}{\Gamma(a)\Gamma(b)}F(c-a, c-b, c-a-b+1, 1-z).
\eea
Since $\beta \leq 0$, the second term vanishes automatically at $z=1$. The quasinormal modes
are then given by (\ref{qnm}), yielding
\bea
\omega_{S} &=& \pm \frac{\xi_{S}}{l} - \frac{2i}{l}\left(n + \frac{d-3}{4}\right),\non\\
\omega_{V} &=& \pm \frac{\xi_{V}}{l} - \frac{2i}{l}\left(n + \frac{d-1}{4}\right),\non\\
\omega_{T} &=& \pm \frac{\xi_{T}}{l} - \frac{2i}{l}\left(n + \frac{d+1}{4}\right),
\label{qnm2}
\eea
where $\xi^{2}_{\mathrm{S}} = Q_{\mathrm{S}} - \left(\frac{d-3}{2}\right)^{2}$, and similarly for
the vector and tensor modes. In general, one notes that the three types of perturbation give
distinct quasinormal frequencies.

Turning now to odd dimensions, let us first consider the case of $d=5$. The subtlety here is that
$c-a-b$ is an integer, and the analytic continuation to $z=1$ contains logarithmically divergent terms.
For the scalar perturbation, we have $\beta_{S} = 1/4$ and $c-a-b = 0$. The master field
near $z=1$ is then given by \cite{Abram}
\bea
\Phi &=& z^{\alpha}(1-z)^{1/4} \frac{\Gamma(a+b)}{\Gamma(a)\Gamma(b)}\sum_{n=0}^{\infty}
\frac{(a)_{n}(b)_{n}}{(n!)^{2}}[2\psi(n+1)
- \psi(a+n) \non\\
&-&\psi(b+n) - {\mathrm{ln}}(1-z)](1-z)^{n},\non\\
\eea
where $(a)_{n} = \Gamma(a+n)/\Gamma(a)$, and $\psi(z) = \Gamma^{\prime}(z)/\Gamma(z)$.
Vanishing of $\Phi$ at infinity is guaranteed by choosing $a = -n$ or $b=-n$. However, since,
$c-a-b = 0$, this can be re-written as (\ref{qnm}), namely $c-a = -n$ or $c-b = -n$.
The scalar quasinormal modes are
\bea
\omega_{\mathrm{S}} = \pm \frac{\xi_{\mathrm{S}}}{l} -\frac{2i}{l}\left(n + \frac{1}{2}\right).
\eea
For the vector modes, we have $\beta_{V} = -1/4$, and $c-a-b = 1$.
The master field is then given by
\bea
\Phi =&=& z^{\alpha}(1-z)^{-1/4}F(a,b,a+b+1,z),
\eea
where, for $(m=1,2,3,...)$, we have
\bea
F(a,b,a+b+m,z) &=& \frac{\Gamma(m)\Gamma(a+b+m)}{\Gamma(a+m)\Gamma(b+m)}\sum_{n=0}^{m-1}
\frac{(a)_{n}(b)_{n}}{n!(1-m)_{n}}(1-z)^{n}\non\\
&-&
\frac{\Gamma(a+b+m)}{\Gamma(a)\Gamma(b)}(z-1)^{m}\sum_{n=0}^{\infty}\frac{(a+m)_{n}(b+m)_{n}}{n!(n+m)!}
(1-z)^{n}[{\mathrm{ln}}(1-z) \non\\
&-& \psi(n+1) -\psi(n+m+1) + \psi(a+n+m) + \psi(b+n+m)].
\label{5d}
\eea
The quasinormal modes are now given by $a+1 = -n$ or $b+1 = -n$. Note that these conditions
also ensure the vanishing of the logarithmic terms in (\ref{5d}). Equivalently, we can write
$c-a = -n$ or $c-b = -n$, yielding
\bea
\omega_{\mathrm{V}} = \pm \frac{\xi_{\mathrm{V}}}{l} -\frac{2i}{l}\left(n + 1\right).
\eea
Finally, the tensor modes have $\beta_{\mathrm{T}} = -3/4$, with $c-a-b = 2$. The Dirichlet
boundary condition on $\Phi$ is enforced by
setting $a+2 = -n$ or $b+2 =-n$ (equivalently $c-a = -n$ or $c-b = -n$). This yields the quasinormal
modes
\bea
\omega_{\mathrm{T}} = \pm \frac{\xi_{\mathrm{T}}}{l} -\frac{2i}{l}\left(n + \frac{3}{2}\right).
\eea

The generalization to all odd dimensions follows suit, with the knowledge that $\beta \leq 0$
for all perturbations, and $c-a-b = m$, with $(m = 1,2,3,...)$.
The quasinormal modes can again be written in the form (\ref{qnm2}).

\section{Discussion}
The determination of quasinormal modes is a subject of general interest in gravitational physics.
These modes describe the response of a black hole to a small perturbation, and in general
they depend only on the parameters of the black hole as well as the parameters of the field probe.
Since the introduction of the proposed correspondence between anti-de Sitter gravity and boundary conformal field theory,
much effort has been devoted to the computation of quasinormal modes of anti-de Sitter black holes.
In most instances, these modes  can only be determined numerically due to the difficulty
of solving the master equations analytically. However, as already pointed out,
the three-dimensional BTZ black hole provides a very useful example where
the quasinormal modes can be determined exactly. This allows one to provide a non-trivial check of
the AdS/CFT correspondence. It was shown in \cite{BSS}, that the quasinormal modes
of the BTZ black hole are identical to the momentum space poles of the retarded correlator
in the strongly coupled dual conformal field theory.

Exactly solvable models are always useful in either gravity or quantum field theory. Thus,
our goal in this paper has been to present an example of a black hole in four and higher dimensions,
for which
the gravitational quasinormal modes can be computed exactly.
We have identified the massless anti-de Sitter black hole with negative curvature horizon as
such an example.
Using the formalism of Kodama and Ishibashi, we have considered the gravitational perturbations
of the massless topological black hole in all dimensions. We have shown that the master equations
for the gravitational modes are identical to the equation for a massive scalar field, for various
values of the scalar mass parameter. This wave equation
can be solved exactly in terms of hypergeometric functions in all cases. The gravitational quasinormal modes
have been defined by requiring the
solution to be ingoing at the horizon
and vanishing at infinity. These boundary conditions are consistent with
the AdS/CFT correspondence \cite{Keski,Starinets1,Kovtun}. Imposing these boundary conditions led us to an exact
determination of all quasinormal modes.
This result represents the first exact analytic determination of gravitational quasinormal modes in four and higher
dimensions. We have seen that scalar, vector, and tensor modes are distinct in general.
Furthermore, the dominant late time
decay behaviour is given by the mode with smallest imaginary part, namely the $n=0$ scalar
frequency.

From the point of view of the AdS/CFT correspondence, it is useful to express the quasinormal modes in terms of the
Hawking temperature $T_{H}$ of the conformal field theory, and the conformal weights of the fields.
Recall that the conformal weight of a massive scalar field in d-dimensional anti-de Sitter space is given by
\bea
\Delta = \frac{1}{2}\left((d-1) + \sqrt{(d-1)^{2} + 4m^{2}l^{2}}\right).
\label{weight}
\eea
Since the gravitational master fields correspond to scalar fields with masses given by eqn. (\ref{masses}),
we can assign a conformal weight to each master field according to (\ref{weight}). For dimension greater than four,
we have
$\Delta_{\mathrm{S}} = (d-3), \Delta_{\mathrm{V}} = (d-2), \Delta_{\mathrm{T}} = (d-1)$.
The form of the quasinormal modes then takes a unified form as
\bea
\omega_{S} &=& \pm 2 \pi T_{H}\xi_{S} - 4 \pi i T_{H}\left(n + \frac{\Delta_{\mathrm{S}}}{2}
- \left(\frac{d-3}{4}\right)\right),\non\\
\omega_{V} &=& \pm 2 \pi T_{H}\xi_{V} - 4 \pi i T_{H}\left(n + \frac{\Delta_{\mathrm{V}}}{2}
-\left(\frac{d-3}{4}\right)\right),\non\\
\omega_{T} &=& \pm 2 \pi T_{H}\xi_{T} - 4 \pi i T_{H}\left(n + \frac{\Delta_{\mathrm{T}}}{2}
- \left(\frac{d-3}{4}\right)\right).
\label{qnmcft}
\eea
In four dimensions, we note that $\Delta_{\mathrm{S}} = \Delta_{\mathrm{V}} = 2$, and thus
the quasinormal modes (\ref{qnm4d}) can also be written in the form (\ref{qnmcft}).
According to the AdS/CFT correspondence, the prediction for the poles of the retarded
correlation functions of the operators associated to each of the gravitational master fields
is then given by (\ref{qnmcft}).

It is worth observing that the real part of the quasinormal modes is independent of the mode number
$n$, while the imaginary part grows linearly with $n$. In \cite{Starinets},
quasinormal modes for various perturbations of the five-dimensional anti-de Sitter black hole with
flat horizon were computed numerically. The structure present in (\ref{qnmcft}), with linear dependence
on the mode number and conformal weight, is similar to that found in \cite{Starinets}.
Recently, the analytic form of the quasinormal modes of the Schwarzschild black hole has been
determined in the large damping ($n \rightarrow \infty$) limit \cite{Motl1,Motl2}.
It has been shown that the imaginary part also has a linear dependence on the mode number $n$ in
this limit \cite{Motl1,Motl2,Pad,Medved,Schiappa}.

\noindent{\bf Acknowledgements}\\[.5ex]
We would like to thank Dr. A. Milki and Dr. J. Chueh for their expertise. This paper is dedicated
to Sufia \'{A}ine.


\begin{thebibliography}{999}
\bibitem{Z1} F. Zerilli, Phys. Rev. Lett. 24 (1970) 737.
\bibitem{Z2} F. Zerilli, Phys. Rev. D2 (1970) 2141.
\bibitem{RW} T. Regge and J.A. Wheeler, Phys. Rev. 108 (1957) 1063.
\bibitem{Fiziev1} P. Fiziev, Class. Quantum Grav. 23 (2006) 2447; gr-qc/0509123.
\bibitem{Fiziev2} P. Fiziev, {\em On the Exact Solutions of the Regge-Wheeler Equation in the Schwarzschild Black
Hole Interior}, gr-qc/0603003.
\bibitem{Maldacena} J. Maldacena, Adv. Theor. Math. Phys. 2 (1998) 231;
hep-th/9711200.
\bibitem{Polyakov} S.S. Gubser, I.R. Klebanov, and A.M. Polakov, Phys. Lett. B428 (1998)
105; hep-th/9802109.
\bibitem{Witten} E. Witten, Adv. Theor. Math. Phys. 2 (1998) 253; hep-th/9802150.
\bibitem{ChanMann} S.F.J. Chan and R.B. Mann, Phys. Rev. D55 (1997) 7546; gr-qc/9612026.
\bibitem{HH} G.T. Horowitz and V.E. Hubeny, Phys. Rev.
D62 (2000) 024027; hep-th/9909056.
\bibitem{BSS} D. Birmingham, I. Sachs, and S.N. Solodukhin, Phys. Rev. Lett. 88 (2002) 151301; hep-th/0112055.
\bibitem{BTZ} M. Ba{\~n}ados, C. Teitelboim, and J. Zanelli,
Phys. Rev. Lett. 69 (1992) 1849; hep-th/9204099.
\bibitem{Cardoso1} V. Cardoso and J.P.S. Lemos, Phys. Rev. D63 (2001) 124015; gr-qc/0101052.
\bibitem{Bir1} D. Birmingham, Phys. Rev. D64 (2001) 064024; hep-th/0101194.
\bibitem{Keski} E. Keski-Vakkuri, Phys. Rev. D59 (1999) 104001; hep-th/9808037.
\bibitem{Lemos} J.P.S. Lemos, Phys. Lett. B353 (1995) 46; gr-qc/9404041.
\bibitem{Mann} R.B. Mann, Class. Quantum Grav. 14 (1997) L109; gr-qc/9607071.
\bibitem{Vanzo} L. Vanzo, Phys. Rev. D56 (1997) 6475; gr-qc/9705004.
\bibitem{Brill} D.R. Brill, J. Louko, and P. Peld\'{a}n, Phys. Rev. D56  (1997) 3600; gr-qc/9705012.
\bibitem{Bir2} D. Birmingham, Class. Quantum Grav. 16 (1999) 1197; hep-th/9808032.
\bibitem{Aros} R. Aros, C. Mart\'{\i}nez, R. Troncoso, and J. Zanelli, Phys. Rev. D67 (2003) 044014; hep-th/0211024.
\bibitem{Mann2} S.F.J. Chan and R.B. Mann, Phys. Rev. D59 (1999) 064025.
\bibitem{Mann3} E. Abdalla, R.B. Mann, and B. Wang, Phys. Rev. D65 (2002) 084006; hep-th/0107243.
\bibitem{Pap} G. Koutsoumbas, S. Musiri, E. Papantonopoulos, and G. Siopsis,
{\em Quasinormal Modes of Electromagnetic Perturbations of Four-Dimemsional Topological Black
Holes with Scalar Hair}, hep-th/0606096.
\bibitem{Kodama1} H. Kodama and A. Ishibashi, Prog. Theor. Phys. 110 (2003)701;
hep-th/0305147.
\bibitem{Cardoso2} V. Cardoso and J.P.S. Lemos, Phys. Rev. D64 (2001) 084017; gr-qc/0105103.
\bibitem{Cardoso3} V. Cardoso and J.P.S. Lemos, Class. Quantum Grav. 18 (2001) 5257; gr-qc/0107098.
\bibitem{Kodama2} H. Kodama and A. Ishibashi, Prog. Theor. Phys. 110 (2003) 901; hep-th/0305185.
\bibitem{Gibbons} G. Gibbons and S.A. Hartnoll, Phys. Rev. D66 (2002) 064024; hep-th/0206202.
\bibitem{BF1} P. Breitenlohner and D.Z. Freedman, Phys. Lett. B115 (1982) 197.
\bibitem{BF2} P. Breitenlohner and D.Z.  Freedman, Ann. Phys. (NY) 144 (1982) 249.
\bibitem{Starinets1} D.T. Son and A.O. Starinets, JHEP 0209 (2002) 042; hep-th/0205051.
\bibitem{Kovtun} P.K. Kovtun and A.O. Starinets, Phys. Rev. D72 (2005) 086009; hep-th/0506184.
\bibitem{Konoplya} R.A. Konoplya, Phys. Rev. D68 (2003) 124017; hep-th/0309030.
\bibitem{Schiappa} J. Natario and R. Schiappa, Adv. Theor. Math. Phys. 8 (2004) 1001; hep-th/0411267.
\bibitem{Abram} M. Abramowitz and I.A. Stegun, {\em Handbook of
Mathematical Functions}, Dover, New York, 1970.
\bibitem{Starinets} A. N\'{u}\~{n}ez and A.O. Starinets, Phys. Rev. D67 (2003) 124013; hep-th/0302026.
\bibitem{Motl1} L. Motl, Adv. Theor. Math. Phys. 6 (2003) 1135; gr-qc/0212096.
\bibitem{Motl2} L. Motl and A. Neitzke, Adv. Theor. Math. Phys. 7 (2003) 307; hep-th/0301173.
\bibitem{Pad} T. Padmanabhan, Class. Quantum Grav. 21 (2004) L1; gr-qc/0310027.
\bibitem{Medved} A.J.M. Medved, D. Martin, and M. Visser, Class. Quantum Grav. 21 (2004) 2393;
gr-qc/0310097.
\end{thebibliography}
\end{document}